# Nonresponse Bias Analysis in Longitudinal Studies: A Comparative Review with an Application to the Early Childhood Longitudinal Study

**Corresponding author:** Yajuan Si, Research Associate Professor,
Survey Research Center, Institute for Social Research, University of Michigan, Ann Arbor
ISR 4014, 426 Thompson St, Ann Arbor, MI 48104
Email: yajuan@umich.edu; Phone: 734-7646935; Fax: 734-7648263

YAJUAN SI is a Research Associate Professor in the Survey Research Center of the Institute for Social Research at the University of Michigan, ISR 4014, 426 Thompson St, Ann Arbor, MI 48104; Email: yajuan@umich.edu.

RODERICK J.A. LITTLE is the Richard D. Remington Distinguished University Professor in the Department of Biostatistics, Research Professor in the Institute for Social Research, and Professor in the Department of Statistics at the University of Michigan, M4071 SPH II, 1415 Washington Heights, Ann Arbor, Michigan 48109; Email: rlittle@umich.edu.

YA MO is an Assistant Professor of Curriculum, Instruction, and Foundational Studies at Boise State University and a research fellow at the National Institute of Statistical Sciences. 1750 K Street, NW, Suite 1100, Washington, DC 20006; Email: ymo@niss.org.

NELL SEDRANSK is the Director Emeritus of the DC Office of the National Institute of Statistical Sciences and Professor of Statistics at North Carolina State University. 1750 K Street, NW, Suite 1100, Washington, DC 20006; Email: NSedransk@niss.org.

# Nonresponse Bias Analysis in Longitudinal Studies: A Comparative Review with an Application to the Early Childhood Longitudinal Study

*Abstract*

Longitudinal studies are subject to nonresponse when individuals fail to provide data for entire waves or particular questions of the survey. We compare approaches to nonresponse bias analysis (NRBA) in longitudinal studies and illustrate them on the Early Childhood Longitudinal Study, Kindergarten Class of 2010-11 (ECLS-K:2011). Wave nonresponse with attrition often yields a monotone missingness pattern, and the missingness mechanism can be missing at random (MAR) or missing not at random (MNAR). We discuss weighting, multiple imputation (MI), incomplete data modeling, and Bayesian approaches to NRBA for monotone patterns. Weighting adjustments can be effective when the constructed weights are correlated with the survey outcome of interest. MI allows for variables with missing values to be included in the imputation model, yielding potentially less biased and more efficient estimates. We add offsets in the MAR results to provide sensitivity analyses to assess MNAR deviations. We conduct NRBA for descriptive summaries and analytic model estimates in the ECLS-K:2011 application. The strength of evidence about our NRBA depends on the strength of the relationship between the fully observed variables and the key survey outcomes, so the key to a successful NRBA is to include strong predictors.

*1. Introduction*

Nonresponse bias analysis has become critical to data quality assessment in sample surveys because of rapidly decreasing response rates (Brick & Williams, 2013; Büttner et al., 2021; de Leeuw, Hox, and Luiten, 2018; Groves, 2006; Hedlin, 2020). Cross-sectional surveys are subject to *unit* nonresponse, where individuals fail to respond to the survey because of noncontact or refusal, and *item* nonresponse, where individuals fail to respond to particular questions. In longitudinal panel surveys, an additional source of missing data is *wave* nonresponse, where some individuals fail to provide data for entire waves of the survey, because they are out of contact or not available to interview, attrition, or refusal. Survey variables are the administered measures only collected for respondents to the study. The analysis of survey variables models survey outcomes with correlated predictors. Nonresponse bias depends on the specific analysis and the difference between respondents and nonrespondents. Nonresponse bias analysis (NRBA) requires information available for both respondents and nonrespondents, such as survey measures in previous waves before dropout and auxiliary variables in the sampling frame or from external data sources.

This work responds to a request from the U.S. National Center for Education Statistics (NCES) to create exemplars of NRBA to guide survey analysts. NCES requires NRBA to be conducted if the response rate falls below 85%. Si et al. (2022a) provided a ten-step exemplar for addressing NRBA in *cross-sectional* surveys:

1) Analyze missing-data patterns.

2) Identify key survey variables and associated analyses.

3) Model key survey variables as a function of fully observed predictors.

4) Seek strong observed predictors in auxiliary data.

5) Model unit nonresponse as a function of observed predictors.

6) Assess observed predictors for the potential for bias adjustment.

7) Assess the effects of nonresponse weighting adjustments on key survey estimates.

8) Compare the survey with external data using summary estimates of key survey variables.

9) Perform a sensitivity analysis to assess the impacts of deviations from the assumption of missing at random (MAR).

10) Conduct item nonresponse bias analyses for all analyzed variables.

In this article, we discuss extensions of the ten-step approach for an NRBA of *longitudinal* data and conduct a comparative review of existing literature work to handle nonresponse in longitudinal studies. In some respects, missing data in longitudinal and cross-sectional surveys are similar; if the data from different waves of the survey are concatenated into one row for each survey unit, with a set of columns for each of the waves – sometimes called the "wide view" of the data – then the missing data create a multivariate pattern of missing data, similar to the pattern for cross-sectional data with unit and item nonresponse. The main differences are that the number of variables may be much greater when variables are repeatedly measured across waves. A distinguishing feature is the fitting of longitudinal models, such as growth mixture models to the "long view" of the data, or a vertical data structure, where each survey unit has multiple rows corresponding to different waves (Fitzmaurice et al., 2012; Muthén, 2004).

A common approach to missing data is complete-case analysis (CCA), where analysis is restricted to individuals who respond to all the waves of the survey. This approach does not exploit information in incomplete cases and can result in bias when the reasons for missing data are associated with the values of survey measures. Available-case analysis (ACA) includes the data that are complete on the set of variables in analysis (Little & Rubin, 2019). Our NRBA compares inferences for questions of interest based on alternative approaches to adjusting for nonresponse, where CCA includes different weighting methods and ACA includes weighting, multiple imputation (MI; Rubin, 1987), multilevel models with maximum likelihood (ML) estimation (e.g., Fitzmaurice et al., 2009), marginal models fitted by generalized estimating equations (GEE; Liang & Zeger, 1986), and Bayesian modeling (Daniels & Hogan, 2008). Furthermore, we conduct sensitivity analysis to assess the impact of deviations from MAR, an assumption that is often required by standard approaches. The basic premise is that if inferences are similar under the alternative approaches, then nonresponse bias is not a major concern.

A critical feature is to characterize the strength of evidence about nonresponse bias, based on the strength of the relationship between the characteristics in the nonresponse adjustment and the key survey outcomes of interest (Little et al., 2022; Little & Vartivarian, 2005), as contained in Steps 2) – 7) in the cross-sectional NRBA. Extending to longitudinal NRBA, we summarize and re-organize the ten steps into five major steps, where one major step in longitudinal NRBA may cover multiple steps in cross-sectional NRBA, to focus on the comparison of different adjustment methods. We provide a comparative review and summarize our NRBA with longitudinal data in the following steps.

1) Identify key survey variables and associated analyses. Model key survey variables as a function of fully observed predictors. Seek strong observed predictors in auxiliary data.

2) Analyze missing-data patterns, including unit, wave, and item nonresponse.

3) Model unit nonresponse as a function of observed predictors. If available, strong predictors observed for both respondents and nonrespondents should be included as covariates in the analysis.

4) Comparison of bias adjustment approaches.

   a) weighting; b) multiple imputation; c) incomplete data modeling; d) Bayesian methods

If available, external data are leveraged for comparison using summary estimates of key survey variables.

5) Perform a sensitivity analysis to assess the impacts of deviations from the MAR assumption.

We apply methods to data from the Early Childhood Longitudinal Study, Kindergarten Class of 2010–11 (ECLS-K:2011; https://nces.ed.gov/ecls/kindergarten2011.asp). The ECLS-K:2011 followed a cohort of children from their kindergarten year (the 2010–11 school year, referred to as the base year, 5-6 year old) through the 2015–16 school year, when most of the children were 10-11 year old in the fifth grade. The ECLS-K:2011 study involves a multi-stage probability design, sequentially selecting geographic areas (counties or groups of contiguous counties) as primary sampling units, schools, and children to collect national data on children's home and school experiences, growth, and learning. During the 2010–11 school year, when both a fall and a spring data collection were conducted, approximately 18,170 kindergartners from about 1,310 schools and their parents, teachers, school administrators, and before- and after-school care providers participated in the study. We focus on the kindergartner respondents at the base year and examine their follow-up participation in the study. Fall and spring data collections were also conducted during the first- and second-grade years, with the fall collection restricted to one-third of the sample selected for the study. In the third, fourth, and fifth grades, a spring data collection was conducted for the entire sample of children who participated in previous waves. The study followed students who moved away from their original base-year schools after the spring base-year data collection, to their new schools. Some movers were followed with certainty, and others were subsampled. The sample sizes in the follow-up waves decreased because of the mover subsampling and attrition, which could also result in systematic differences in children assessment outcomes or developmental trajectories between the study participants and the target population. It is possible that the underlying missingness mechanisms for movers and attriters are different. Nevertheless, we do not have enough information to separate movers from attriters and have to treat all the nonparticipants to a wave as one type of nonresponse, and the extension to handle multiple mechanisms will be discussed in Section 5. We apply an NRBA to evaluate the impact of nonresponse on child assessments in the analysis of the longitudinal ECLS-K:2011 study.

We focus on descriptive summaries and mixed-effect model regression estimates. Estimates of subgroup means, such as average assessment scores across race/ethnicity and income groups, are of interest to policymakers. Education researchers are also interested in estimating growth curves representing trajectories of child development across time while considering the correlation between repeated measurements.

There is a rich literature on missing data in longitudinal studies (e.g., Zhou & Kim, 2012). Our paper differs from previous review articles (e.g., Brick, 2013; Ibrahim et al., 2005; Ibrahim & Molenberghs, 2009) by simultaneously handling unit, wave, and item nonresponse in a longitudinal survey, attrition-adjusted weighting, extensive discussions of MI related developments, and proposed extensions of sensitivity analysis. The paper structure is as follows. Section 2 introduces the background with missing longitudinal data. Section 3 describes NRBA methods for a monotone missingness pattern with detailed comparisons between weighting, MI, incomplete data modeling, and sensitivity analysis. Section 4 illustrates the discussed methods with the ECLS-K: 2011 study. We conclude recommendations and discussions in Section 5.

## 2. Background

Suppose the study collects baseline measures $(X_0, Y_0)$ and $T$ follow-up waves of the survey variables $(X_t, Y_t)$, for $t = 1, \ldots, T$, where the analysis of interest focuses on the descriptive summaries of the outcome $Y$ and the regression of $Y$ on the covariates $X$ across $T$ waves. In addition to the time-varying covariates $X$, time-invariant covariates $Z$ are also available, such as the base-year sample design features, survey administrative information, and other variables collected at baseline that do not change over time.

A complete case has all the variables $(Z, X_0, X_1, \ldots, X_T, Y_0, Y_1, \ldots, Y_T)$ observed. Individuals dropping out between waves $t - 1$ and $t$ have $(Z, X_0, X_1, \ldots, X_{t-1}, Y_0, Y_1, \ldots, Y_{t-1})$ observed but $(X_t, \ldots, X_T, Y_t, \ldots, Y_T)$ missing. For participants who miss a wave and return to the study later, the missingness is intermittent. Along with unit and wave nonresponses, item nonresponse arises when participants only provide answers to a subset of the survey variables $(X_0, X_1, \ldots, X_T, Y_0, Y_1, \ldots, Y_T)$. The information in $Z$ is often fully observed.

The missingness mechanism refers to the relationship between missingness and the variables in the analysis and plays a crucial role in NRBA. Rubin (1976) treated the response indicators $R$ as random variables and characterized the missingness mechanism by the conditional distribution of $R$ given the data. If the response indicator $R$ does not depend on any values of the data, the missingness is missing completely at random (MCAR). MCAR is a strong and often implausible assumption. A weaker assumption is MAR, where the missingness depends only on values of observed variables. If missingness depends on missing variables after conditioning on the observed variables, the data are called missing not at random (MNAR). The effectiveness of NRBA depends on the quantity of interest and whether the missingness mechanism is ignorable given observed information, i.e., MAR and distinct model parameters for the analysis and missingness mechanism. The MAR assumption is untestable from the collected data alone without additional structural assumptions, so in our NRBA we assess the sensitivity of inferences to deviations from MAR.

In many longitudinal studies, including the ECLS-K:2011 study, the pattern of missing data is predominantly monotone, because individuals dropping out at wave $t$ do not reenter the study in a later wave. Accordingly, we focus here on methods suitable for a monotone pattern. The extensions to non-monotone dropouts are discussed in Section 5. Let $R_t$ denote the wave or unit response indicator at wave $t$, where $R_t = 1$ if the individual responds at wave $t$, otherwise $R_t = 0$ if the individual drops at wave $t$. For the monotone pattern, the response indicators in later waves will all be 0, i.e., $R_{t+1} = \cdots = R_T = 0$ if $R_t = 0$, and we have $\Pr(R_t = 0) = \Pr(R_t = R_{t+1} = \cdots = R_T = 0)$ and $\Pr(R_t = 1) = \Pr(R_t = R_{t-1} = \cdots = R_1 = 1)$.

Even with $R_t = 1$, the participants may only provide answers to a subset of the survey variables, where $(X_t, Y_t)$ is subject to item nonresponse. Theoretically, we can introduce response indicators to represent different item missingness mechanisms. However, since item nonresponse depends on the specific measure and wave, with many survey variables analyzed across multiple time points, modeling of item response indicators will become complex and difficult for estimation. Therefore, we usually assume the item nonresponse is MAR, i.e., ignorable. An analysis of nonignorable item nonresponse often considers one variable of interest across waves or multivariate items in a cross-sectional study, discarding unit nonrespondents (e.g., Ibrahim et al., 2005). In the ECLS-K:2011 study, the proportions of item nonresponse are low (detailed values are presented below). Accordingly, we assume ignorable item nonresponse and potentially nonignorable unit or wave nonresponse through a joint model $(R_t, X_t, Y_t)$. The setting can be extended to multivariate response indicators with further dimension reduction techniques.

### 3. *NRBA methods for a monotone missingness pattern.*

We outline the NRBA steps with longitudinal data and compare different methods in detail.

1) Identify key survey variables and associated analyses.

For the survey outcome measured for individual $i$ at wave $t$, $Y_{it}$, we consider an analysis model that includes time-varying covariates $X_{it}$ and time-invariant covariates $Z_i$ and specify the mean structure as

$$E(Y_{it}) = \alpha_0 + \vec{\beta}_1 Z_i + \vec{\beta}_2 X_{it} + \vec{\beta}_3 Z_i X_{it}, \qquad (1)$$

for $i = 1, \dots, n$, and $t = 0, 1, \dots, T$. The quantities of interest include the descriptive summaries of $Y_{it}$ and the model coefficients $(\vec{\beta}_1, \vec{\beta}_2, \vec{\beta}_3)$. In the ECLS-K:2011 study, we model the reading assessment outcome $(Y_{it})$ with selected time-varying child, family, and school characteristics and wave variables $(X_{it})$ across race/ethnicity categories $(Z_i)$.

The correlation of the repeated measures across $T$ time points for individual $i$ needs to be accounted for in the variance-covariance specification, through child-level random effects or a marginal correlation structure. The clustering of children, such as schools and geographic areas, also needs to be modeled.

2) Analyze missing-data patterns, including unit, wave, and item nonresponse.

In addition to the monotone dropout pattern for individuals, the survey outcomes and covariates are subject to item nonresponse. Methods for addressing this make different assumptions about the missing data mechanisms.

3) Model unit nonresponse as a function of observed predictors.

Little & Vartivarian (2005) conclude that useful variables in the NRBA are strongly correlated with the outcome (primary) and the response indicator (secondary) and have to be available for both respondents and nonrespondents. The design features of the ECLS-K:2011 survey, including primary sampling units and survey weights should be included in the NRBA. To quantify the correlation, Collins et al. (2001) and Little et al. (2022) classify variables' strength of relationship with the survey outcome and response indicator and their effect on bias and variance of MI and inverse propensity weighting for estimates of means, compared to unadjusted analyses based on the complete cases. NRBA is only effective when the auxiliary variables are related to the survey outcome variables.

4) Comparison of bias adjustment approaches.

The bias depends on the form of the analysis model. A default approach for the analysis of longitudinal data with missing data is unadjusted CCA, where cases that are not complete over all waves are dropped, and the analysis of complete cases may include sampling weights but does not adjust for nonresponse arising from attrition after baseline. This analysis yields consistent estimates if missingness depends on fully observed baseline variables, but conditional on those variables, does not depend on repeated measures post-baseline (Little & Rubin, 2019).

The main approaches to reducing nonresponse bias from this default CCA are (a) to estimate nonresponse weights that multiply the sampling weights in CCA; or (b) to impute the missing values, using MI combining rules to propagate imputation error in the inferences. We can also fit a model to the complete and incomplete data, i.e., available data, using a method that allows for missing data in the repeated measures, such as MLE, GEE, or Bayesian estimation. Weighting and imputation can also be applied to ACA, which includes the set of cases that are complete on the set of variables included in the particular analysis of interest. In Sections 3.1-3.3 we describe the main features of each of these approaches, assuming missingness is MAR or MNAR. In Section 3.4 we describe a sensitivity analysis to assess deviations from MAR.

*3.1 Weighting*

Weighting is a common method of unit/wave nonresponse adjustment, with weights defined as the inverse of estimated response propensities (Brick, 2013; Brick & Montaquila, 2009; Haziza & Lesage, 2016; Kalton & Flores-Cervantes, 2003). These can be estimated as the inverse of the response rates in adjustment cells formed from variables observed for respondents and nonrespondents, also known as weighting classes, or more generally by regressing response indicators on covariates, and creating weights that are the inverse of the estimated response probabilities (Chen et al., 2015). In longitudinal studies, different weights can be created depending on the set of complete cases being analyzed (Lynn & Watson, 2021; Si et al., 2022b).

For example, in the ECLS-K:2011 study, different weights are constructed based on the module questions answered by children, parents, teachers, and schools.

Weighting usually assumes MAR and assumes a non-zero response probability for each observation. In ACA, weights at baseline or in the previous wave can be applied to adjust for nonresponse across waves, and every case is assigned a weight. Suppose we are constructing weights for wave $t$ and assume MAR applies to all nonresponse mechanisms at each wave. A standard set of weights based on the baseline covariates are inverses of the estimated response propensity given the baseline information,

$$\hat{w}_t^{base} = Pr^{-1}(R_t = 1|Z, X_0, Y_0). \qquad (2)$$

An alternative approach, which makes better use of available data, expresses the response probability as a product of conditional probabilities given response at previous waves (Little & David, 1983). These conditional probabilities can then condition on all available information in earlier waves. That is,

$$\begin{aligned}\hat{w}_t^{seq} &= Pr^{-1}(R_t = 1|Z, X_0, Y_0, X_1, Y_1, \ldots, X_{t-1}, Y_{t-1}) \qquad (3)\\ &= Pr^{-1}(R_t = R_{t-1} = \cdots = R_1 = 1|Z, X_0, Y_0, X_1, Y_1, \ldots, X_{t-1}, Y_{t-1})\\ &= Pr^{-1}(R_1 = 1|Z, X_0, Y_0)Pr^{-1}(R_2 = 1|R_1 = 1, Z, X_0, Y_0, X_1, Y_1) \ldots\\ &\quad \ldots Pr^{-1}(R_t = 1|R_1 = 1, \ldots, R_{t-1} = 1, Z, X_0, Y_0, X_1, Y_1, \ldots, X_{t-1}, Y_{t-1}).\end{aligned}$$

Here we have

$$\hat{w}_t^{seq} = \hat{w}_{t-1}^{seq} * Pr^{-1}(R_t = 1|R_1 = 1, \ldots, R_{t-1} = 1, Z, X_0, Y_0, X_1, Y_1, \ldots, X_{t-1}, Y_{t-1}).$$

In practice, the weights $\hat{w}_t^{base}$ or $\hat{w}_t^{seq}$ are scaled to sum to the available sample size at wave $t$. Kaminska (2022) compares similar weighting methods to $\hat{w}_t^{base}$ and $\hat{w}_t^{seq}$ with rotating panels.

Approaches to modeling the response propensities include parametric models such as logistic regressions and machine learning algorithms such as tree-based methods, boosting methods, and neural networks (e.g., Breiman et al., 1984; Buskirk & Kolenikov, 2015; Kern et al., 2023; Lohr et al., 2015; Sánchez-Cantalejo et al. 2021). With MNAR, strong parametric assumptions must be introduced for the response model estimation (Scharfstein et al., 1999).

With the constructed weights for all available cases, population means are estimated by weighted sample means, regression coefficients from weighted regression models, and longitudinal model parameters estimated by weighted GEE (Robins et al., 1995) or weighted multilevel models (Pfeffermann et al., 1998; Rabe-Hesketh & Skrondal, 2006), where appropriate multiple-level weights could be informative, such as school weights, conditional individual weights and observation weights. Sampling variability can be accounted for by using large sample asymptotics with Taylor series approximation or resampling approaches, such as jackknife and bootstrap replication (e.g., Natarajan et al., 2008). With response propensity score modeling, ideally the uncertainty of weighting construction based on prediction should be also captured with replication methods (e.g., Valliant et al. 2018; Wager et al., 2014).

Weight construction requires that covariates used in the response propensity modeling are fully observed; however, in practice both the outcome and covariates are subject to item nonresponse. Imputation of missing data allows these variables to be included in the weights and is an attractive alternative.

### 3.2 Multiple imputation

To propagate the uncertainty of the imputation model, MI can be applied to simultaneously handle unit and item nonresponse (Rubin, 1987; He et al., 2021; Si, 2012). MI applies either joint or sequential conditional imputation models to generate multiple completed datasets and use MI combining rules for inferences about parameters or population quantities (e.g., Raghunathan et al., 2001; Si & Reiter, 2013). To achieve inferential validity, the specified imputation models should yield plausible imputations, and the imputation model and the analysis model need to be congenial (Xie & Meng, 2017). The imputation model can include auxiliary variables not included in the analysis model of interest, or variables available for imputation but not available in public use files for confidentiality concerns. The relationship between the survey outcome and covariates in the imputation model should be compatible with the substantive model (Bartlett et al., 2014; Enders et al., 2020; Grund et al., 2016; Seaman et al., 2012). With ACA, the missing items in the outcome or the covariates that are time-varying or invariant can be imputed. MI can handle incomplete longitudinal or multilevel data with multivariate mixed-type variables (Audigier et al., 2018; Cao et al., 2022; Demirtas, 2009; Graham & Hofer, 2000; Grund et al., 2019; Van Buuren, 2011; Yucel, 2008).

With a monotone dropout pattern in longitudinal studies, we can sequentially impute missing responses from the first follow-up to the last wave, i.e., working with the wide data format where variables from each wave are concatenated into a single row for each sample unit, where iteration is not needed for monotone data:

$$f(X_1, Y_1 | Z, X_0, Y_0) \to f(X_2, Y_2 | Z, X_0, Y_0, \hat{X}_1, \hat{Y}_1) \to \cdots \to \quad (4)$$
$$f(X_T, Y_T | Z, X_0, Y_0, \hat{X}_1, \hat{Y}_1, \ldots, \hat{X}_{T-1}, \hat{Y}_{T-1}).$$

The sequential imputation approach is attractive in practice to accommodate the ongoing data collection across time, where new wave data are imputed based on historical records and will then be used as covariates for future waves. Joint imputation with wide data is challenging due to a potentially large number of variables to be imputed.

An alternative imputation strategy is to fit a multilevel model with individuals as clusters and repeated measurements as units with the long data format. A two-level imputation approach includes the analysis model ($Y_{it} | X_{it}, Z_i$) as part of the imputation model specification and the marginal imputation for $X_{it}$, to be compatible with the substantive analysis model. The imputation model explicitly accounts for the correlation between repeated measures of the same individual over time by introducing random effects across individuals, e.g., random intercepts and slopes. The multilevel MI is available in the R packages, such as *mice* (Van Buuren, 2012), *jomo* (Quartagno et al., 2019), and *mdmb* (Robitzsch & Lüdtke, 2021). The analysis model could have interactions and non-linear functions of the covariates, or have complex specifications that are not available in the current software implementations, where advanced computational algorithms, such as importance sampling or other rejection algorithms, become useful (Enders et al., 2020; Lüdtke et al., 2019; Quartagno & Carpenter, 2022).

    MI can handle MAR and MNAR mechanisms (e.g., Si et al., 2015, 2016, 2020). The imputation model can also accommodate response indicators and jointly model the missing values and missingness mechanisms $f(X_t, Y_t, R_t)$ via selection models $f(R_t|X_t, Y_t)f(X_t, Y_t)$ (Kenward, 1998), pattern mixture models $f(X_t, Y_t|R_t)f(R_t)$ (Kenward et al., 2003; Little, 1993; Little, 1994; Little & Wang, 1996), and shared parameters $f(R_t|b_t)f(X_t, Y_t|b_t)f(b_t)$, i.e., conditional independence given a shared but unknown quantity $b_t$ (Wu & Carroll, 1988). The selection model $f(R_t|X_t, Y_t)$ depends on the missing outcome, and the parameters may not be identified. The Heckman model, with a probit regression as the selection model, is weakly identified, and identification is strongly dependent on the normality assumptions (Little & Rubin, 2019). Preferred approaches are to make additional assumptions about the form of the mechanism, such as MAR and instrumental variables, or to do a sensitivity analysis by fitting the model for a variety of plausible choices of the sensitivity parameters. Sensitivity analysis with proxy pattern-mixture models can be used to evaluate model robustness against deviations from

MAR (Andridge & Little, 2011, 2020). Deng et al. (2013) and Si et al. (2015, 2016) use additional data from the refreshment samples to assist with the model identification.

*3.3. Repeated measures models that allow for missing data*

An alternative to weighting and MI is to fit a model with the mean structure in (1) to the available data, using a method that allows for missing data in the repeated measures, with parameters estimated by GEE, ML, or Bayesian methods.

GEE introduces a working correlation structure among repeated measures and estimates the marginal mean structure. Applying GEE without nonresponse weights generally yields consistent estimates if missingness does not depend on repeated measures, which is usually a strong assumption. Weighted GEE, including nonresponse weights, can produce valid inferences under weaker MAR assumptions.

ML estimates focus on the analysis model and maximize the incomplete data likelihood. For example, in growth modeling, we obtain ML estimates under the multilevel model specification with random effects with expectation-maximization algorithms under a latent variable modeling framework (Dempster et al., 1977). Structural equation models also treat random effects as continuous latent variables and apply large sample asymptotics under ML, as in the Mplus software (Muthén & Muthén, 2017).

With nonresponse-adjusted weight values for each available case, weighted ML estimates maximize the pseudo-likelihood function, which weights each case's contribution to the likelihood by the inverse of the power of its estimated response propensity. Pfeffermann et al. (1998) and Rabe-Hesketh & Skrondal (2006) decompose the case weights into two levels, where the base weights adjust for the individual response propensity to the baseline study and the conditional follow-up weights adjust for wave nonresponse given the baseline participation.

ML under a repeated-measures model is asymptotically equivalent to MI under that model and hence is likely to achieve similar results unless the sample size is small (Little & Rubin, 2019). The underlying theory of MI is Bayesian, and the inference with combining rules under MI approximates Bayesian posterior computation.

Fully Bayesian methods implement a data augmentation process by iteratively imputing missing values and estimating model parameters conditional on the completed data with a Markov chain Monte Carlo algorithm, where the missing values are drawn from the posterior predictive distributions (Schafer, 1997). With non-informative prior information and large sample sizes, the fully Bayesian estimation should be similar to ML. With informative prior, Bayesian methods will stabilize estimates based on data of small sample sizes. The missing mechanism can be MAR or MNAR. Bayesian selection models (Du et al., 2021) and pattern mixture models (Si et al., 2016) have been developed to jointly model the outcome and response indicator, where the identification requires strong prior information or extra data. The stand-alone software *Blimp* implements Bayesian estimation in a latent variable modeling program and conducts MI (Hayes, 2019; Keller & Enders, 2021). The computation of Bayesian methods for missing data can be complicated with many incomplete variables in large-scale studies, such as our application to the ECLS-K: 2011 data.

### *3.4 Sensitivity analysis*

The methods reviewed in the previous sections generally assume MAR and can handle MNAR but require untestable assumptions to identify parameter estimates. Therefore, we propose and implement a sensitivity analysis to assess the impact of MNAR deviations from MAR. Following the general approach in Giusti & Little (2011), we apply an offset to MAR imputations to model MNAR departures. For example, in an analysis of racial and ethnic differences, we assume:

$$E(Y_{it}|Z_i, X_{it}, R_{it} = 0) = E(Y_{it}|Z_i, X_{it}, R_{it} = 1) + k_t * \sigma_{race[i]}. \tag{5}$$

Here the offset $k_t * \sigma_{race[i]}$ is the product of the pre-specified $k_t$ at the time point $t$ of dropout and the residual standard deviation for respondents stratified by race/ethnicity, where $race[i]$ denotes the race/ethnicity category of the individual $i$. The approach is as follows:

1) Run MAR-based MI.

2) If individual $i$ drops out of the study at time $t$, fit the regression model $(Y_{it}|R_{it} = 1, -)$ conditional on all other available information up to this wave and obtain the residuals.

3) Calculate the residual standard deviation for respondents stratified by race/ethnicity $\sigma_{race[i]}$.

4) With the pre-specified value of $k_t$, add $k_t * \sigma_{race[i]}$ to the MI outputs at time $t$ and obtain the imputed value $\hat{Y}_{it}$.

5) Conditional on the imputed value $\hat{Y}_{it}$, continue the MAR-based MI for $\hat{Y}_{i,t+1}$, and so on.

We try different pre-specified values of $k_t$ as sensitivity analysis. Giusti & Little (2011) consider 0.8, 1.2 and 1.6 to reflect small, medium, and large deviations from MAR, respectively. The selection of pre-specified values is subjective and should depend on the substantive field knowledge.

## 4. Application to the Early Childhood Longitudinal Study

We focus on the spring data collections in the ECLS-K:2011 study and examine the child development from the kindergarten to fifth grade, i.e., with the kindergarten year referred to as the base study ($t = 0$) and five follow-up waves ($t = 1, \ldots, T(= 5)$).

### 4.1 Identify key survey variables and associated analyses

We estimate descriptive summaries and developmental trajectories of the child assessment performances and examine the disparity across sociodemographic subgroups. The assessment outcome is the item response theory (IRT)-based overall scale score for each domain, such as reading and mathematics, specifically the sum of the predicted probabilities that the child would have correctly answered each assessment item. We present estimates of the average assessment scores and compare the wave-specific averages for children of different race/ethnicity. We estimate regression coefficients of interest in a multilevel linear regression model with varying intercepts across children and schools. The mean structure is specified as

$$E(Y_{it}) = \alpha_0 + \beta_1 AGEC_{it} + \beta_2 AGEC_{it}^2 + \beta_3 SEX_i + \vec{\beta}_4 RACETH_i + \vec{\beta}_5 POV_{it} + \\ \beta_6 PNW_{it} + \beta_7 STY_{it} + \vec{\beta}_8 WAVE_{it} + \vec{\beta}_9 WAVE_{it} * RACETH_i. \qquad (6)$$

Here the predictors include the wave indicators $WAVE_{it}$, race/ethnicity indicator $RACETH_i$, their two-way interaction $WAVE_{it} * RACETH_i$, age $AGEC_{it}$ (standardized), the squared term of age $AGEC_{it}^2$, sex $SEX_i$, poverty $POV_{it}$, the proportion of non-white students in the school $PNW_{it}$, and school type $STY_{it}$.

*Table 1. Unit and item nonresponse rates in spring data collections (in percentage).*

|  | Unit | Any item | Per item | |
|---|---|---|---|---|
|  |  |  | Median | Max |
| Kindergarten | - | 5.7 | 0.1 | 4.5 |
| Grade 1 | 11.7 | 6.4 | 1.4 | 4.7 |
| Grade 2 | 18.5 | 6.6 | 2.4 | 5.0 |
| Grade 3 | 23.4 | 7.7 | 3.4 | 6.2 |
| Grade 4 | 27.4 | 8.0 | 4.6 | 7.3 |
| Grade 5 | 30.9 | 10.6 | 5.7 | 8.7 |

*Note: Here the unit nonresponse rates are conditional on the kindergarten response at the base year. We acknowledge the potential shortcoming of not being able to account for the baseline nonresponse, where the base samples could be substantially different from the nonparticipants.*

*4.2 Missingness pattern and mechanism*

The data are subject to unit, wave, and item nonresponse, mainly arising because of attrition from the sample, and the fact that only a subsample of children moving to different schools are followed. The ECLS-K:2011 study defines one student respondent if the student has scoreable reading or mathematics or science data, or at least one executive function score, or a height or weight measurement, or a completed item from the child questionnaire. In our analysis using the reading assessment score as the survey outcome, we treat students with legitimately assigned reading scores as respondents, including missing items.

The collected variables include 88 measures of child-level characteristics, namely sex, age, race/ethnicity, disability and special education status, family-level measures on poverty, socioeconomic and food security status, and school-level characteristics, such as the school type (public or private), region, locality, enrollment, percent of non-white students, lowest and highest
212121

grades offered at the school. We present the summary of unit and item nonresponse in Table 1. The unit nonresponse rates (conditional on the kindergarten response at the base year) increase from 11.7% to 30.9% across waves. The complete cases are individuals who have participated in all follow-up studies and include 9,590 out of the total 14,730 children, which are the available cases who participated in at least one of the follow-up studies. The wave nonresponse pattern is monotone with scattered item nonresponse. We also observe different wave nonresponse rates across subgroups, e.g., across race/ethnicity categories, where the non-Hispanic Black and API children tend to miss the follow-up studies. There are racial differences in the response mechanism, and race is potentially associated with child development, which is the outcome of interest.

The item nonresponse rates of the 88 variables are generally low. Across waves the percentages of respondents with missing values of at least one item are between 5.7% and 10.6%. When we compare item nonresponse rates across measures, the highest nonresponse rate is of the measure in the last wave (8.7%). Considering the proportion of item nonresponse is small and some methods cannot simultaneously handle unit and item nonresponse (e.g., weighting), we address the item nonresponse by imputation and then compare different approaches to handling wave nonresponse.

*4.3 Need for strong predictors observed for both respondents and nonrespondents*

Since the wave nonresponse is monotone, we use collected measures in previous waves as predictors in the NRBA. The list given in Section 4.1 includes all variables in Model (6) and additional measures correlated to the assessment outcome and response indicator.

*Table 2. Missing data adjustment approaches investigated.*

| Method | CC | AC | Assumption | Base weight | Conditional information |
|---|---|---|---|---|---|
| CCA | √ | | MCAR | | - |
| ACA | | √ | MAR | | Available |
| CCA-base-w | √ | | MCAR | √ | - |
| ACA-base-w | | √ | MAR | √ | Available |
| CCA-attr-w | √ | | MAR | √ | Baseline |
| ACA-attr-w | | √ | MAR | √ | Baseline |
| ACA-seq-attr-w | | √ | MAR | √ | Available |
| MI | | √ | MAR | √ | Available |
| ML | | √ | MAR | √ | Available cases and specified models |
| w-ML | | √ | MAR | √ | Available cases and specified models |
| GEE | | √ | MAR | √ | Available cases and specified models |
| w-GEE | | √ | MAR | √ | Available cases and specified models |
| MI-offset | | √ | MNAR | √ | Available |

*Note: CC: complete case; AC: available case; CCA: complete case analysis; ACA: available case analysis; CCA-base-w: base weighted analysis accounting for the cluster structure by individuals with complete cases; ACA-base-w: base weighted analysis accounting for the cluster structure by individuals with available cases; CCA-attr-w: attrition-adjusted weighted analysis accounting for the cluster structure by individuals with complete cases; ACA-attr-w: attrition-adjusted, conditional on the baseline information, weighted analysis accounting for the cluster structure by individuals with available cases; ACA-seq-attr-w: attrition-adjusted, conditional on the previous information, weighted analysis accounting for the cluster structure by individuals with available cases; MI: multiple imputation with sequential wave by wave conditional imputation; ML: maximum likelihood estimates with a specified two-level model for the repeated measures across time; w-ML: weighted maximum likelihood estimates; GEE: generalized estimation equations (GEE) with a specified working correlation matrix; w-GEE: weighted GEE; MI-offset: sensitivity analysis with an offset $k_t \sigma_{race}$ added to the MI-seq imputations, with different pre-specified values $k_t$.*

*4.4 Comparison of bias adjustment methods*

Table 2 summarizes the different NRBA approaches that will be described in detail below. As discussed in Section 3, even though all methods can handle MAR and MNAR unit nonresponse, we have to introduce additional but unverified assumptions for estimation (tailored to the specific analysis of interest) to jointly model the response indicators and survey outcomes or implement sensitivity analysis. In the ECLS-K:2011 application, we conduct NRBA for the general purpose of data quality assessment and illustrate the sensitivity analysis after MI.

The methods of CCA and base weighted CCA assume that attrition is MCAR, and attrition weighted CCA assumes MAR. The ACA, base weighted ACA, attrition weighted ACA in Equation (2), and the sequentially attrition weighted ACA in Equation (3) assume MAR given the available cases. We account for the cluster structure by individuals in weighted analyses. MI refers to the sequential wave by wave conditional imputation, ML fits a two-level model for the repeated measures across time, and w-ML is the weighted maximum likelihood estimate. GEE specifies a first order auto-regressive working correlation matrix, and w-GEE is the weighted GEE with different weight and data choices. Considering MNAR, MI-offset represents the sensitivity analysis with an offset $k_t \sigma_{race}$ added to the MI-seq imputations, with different pre-specified values $k_t$.

*4.4.1 Weighting*

Weighting in practice often assumes MAR and uses observed information to predict response propensity. To construct the attrition-adjusted weights, we fit stepwise (unweighted) logistic regression models of the propensity to respond at each follow-up wave (Little & Vartivarian, 2003), $Pr(R_t = 1|-)$, conditional on baseline information (CCA-attr-w and ACA-attr-w) or all available measures in previous waves (ACA-seq-attr-w). The base weight reflecting the unequal probability of sampling at baseline is included as a candidate in the response propensity models. Section 3.1 gives the details of the constructed weight which is the product of the inverse of the predicted response propensities and the base weight. As an alternative, we use tree-based methods and select the variables and their high-order interactions that are predictive of the response propensity $Pr(R_t = 1|-)$. We apply default values of the complexity parameters in pruning tree size, often resulting in simple trees with few nodes, which can be modified to increase the complexity and prediction power. For our purpose of constructing weights to achieve a balance between bias and variance, we construct the sequentially adjusted weights based on the stepwise logistic regression (ACA-seq-attr-w) and the baseline-adjusted weights (CCA-attr-w and ACA-attr-w) with conditional inference tree methods implemented in the R package *party* (Hothorn et al., 2006, 2021).

The CCA response propensity model shows that race/ethnicity, special education status, assessment outcome in kindergarten, poverty, food security status, school type, enrollment, locale, region, the proportion of non-white and the highest grade in the school affect the tendency to participate in all waves. As a measure of the predictive accuracy of binary responses, the Area Under the Curve (AUC) is 0.64, indicating that the selected covariates have moderate power to predict the response propensity.

For monotone data, ACA is attractive compared to CCA because it makes better use of available information and assumes MAR. The response indicator can use the previous wave or the baseline as the benchmark, yielding sequential and baseline adjustments, respectively.

*Table 3. Summary of the base and attrition-adjusted weights for the five follow-up waves (all with mean =1). The upper bound of efficiency loss due to weighting Loss(w) is defined as $cv^2(w) = var(w)/mean^2(w)$, resulting in the design effect of $1+ cv^2(w)$.*

|  | Min. | 1st Qu. | Median | 3rd Qu. | Max. | SD | Loss(w) | Sample size |
|---|---|---|---|---|---|---|---|---|
| Base-w | 0.05 | 0.69 | 0.93 | 1.17 | 4.51 | 0.53 | 28% | 14730 |
| CCA-attr-w | 0.05 | 0.68 | 0.9 | 1.15 | 7.16 | 0.58 | 33% | 9590 |
| ACA-seq-attr-w1 | 0.05 | 0.69 | 0.93 | 1.17 | 5.31 | 0.54 | 29% | 12840 |
| ACA-seq-attr-w2 | 0.06 | 0.69 | 0.92 | 1.17 | 5.77 | 0.54 | 29% | 11720 |
| ACA-seq-attr-w3 | 0.06 | 0.7 | 0.93 | 1.17 | 5.9 | 0.53 | 28% | 10900 |
| ACA-seq-attr-w4 | 0.06 | 0.7 | 0.93 | 1.17 | 5.7 | 0.53 | 28% | 10200 |
| ACA-seq-attr-w5 | 0.08 | 0.69 | 0.92 | 1.17 | 5.46 | 0.54 | 29% | 9590 |
| ACA-attr-w1 | 0.05 | 0.67 | 0.91 | 1.16 | 8.26 | 0.57 | 33% | 12840 |
| ACA-attr-w2 | 0.05 | 0.67 | 0.92 | 1.17 | 8.34 | 0.56 | 31% | 11720 |
| ACA-attr-w3 | 0.05 | 0.67 | 0.91 | 1.16 | 6.09 | 0.56 | 32% | 10900 |
| ACA-attr-w4 | 0.04 | 0.66 | 0.9 | 1.16 | 29.47 | 0.63 | 40% | 10200 |
| ACA-attr-w5 | 0.04 | 0.65 | 0.88 | 1.16 | 8.45 | 0.61 | 37% | 9590 |

*Note: Base-w: base weight; CCA-attr-w: attrition-adjusted weights accounting for the cluster structure by individuals with complete cases; ACA-attr-w#: attrition-adjusted, conditional on the baseline information, weights accounting for the cluster structure by individuals with available cases in the #th follow-up wave; ACA-seq-attr-w#: attrition-adjusted, conditional on the previous information, weighs accounting for the cluster structure by individuals with available cases in the #th follow-up wave. All reported sample sizes are rounded to the nearest ten due to confidentiality concerns.*
*SOURCE: U.S. Department of Education, National Center for Education Statistics, Early Childhood Longitudinal Study, Kindergarten Class of 2010-11, 2011-16 Spring.*

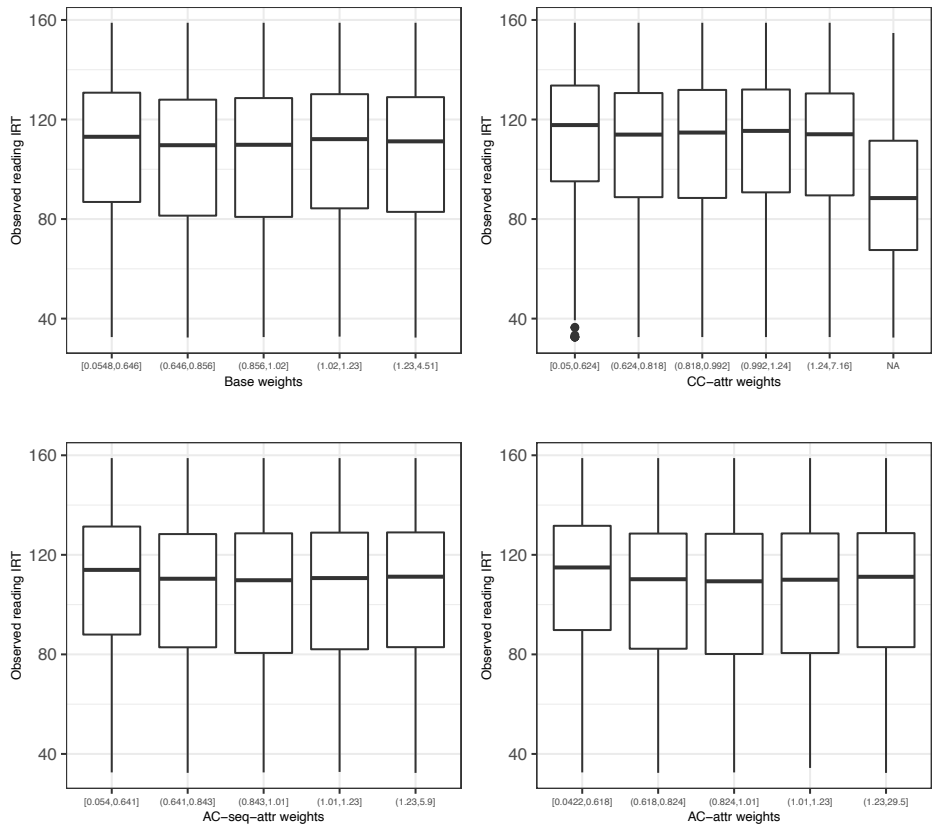

*Figure 1. The distribution of collected reading IRT scores across groups defined by quintiles of different weights. CCA-attr weights: attrition-adjusted weights accounting for the cluster structure by individuals with complete cases; ACA-attr weights: attrition-adjusted, conditional on the baseline information, weights accounting for the cluster structure by individuals with available cases; ACA-seq-attr weights: attrition-adjusted, conditional on the previous information, weighs accounting for the cluster structure by individuals with available cases. SOURCE: U.S. Department of Education, National Center for Education Statistics, Early Childhood Longitudinal Study, Kindergarten Class of 2010-11, 2011-16 Spring.*

The sequential adjustment utilizes analyzed survey variables in the previous waves, which could be highly related to the response mechanism and survey outcome at the current wave and substantially improve the prediction. If the attrition pattern is intermittent, ACA is challenging unless imputation is used to create a monotone pattern; otherwise, CCA can be applied because all conditional information is fully observed. The current weighting practice implemented by the ECLS-K:2011 study adjusts the base-year weight for wave nonresponse. We find that in the logistic regression models for the ACA response indicators, the survey variables in the previous waves are often selected, as well as the base and attrition-adjusted weights. The sequential ACA response propensity models yield five AUC values between 0.59 and 0.66, and the ACA adjustments to the base year have five AUC values around 0.66, indicating similar prediction power. We trimmed the $\hat{w}_3^{seq}$ at wave 3 at its 95% quantile value to avoid an extreme maximum value. Trimming could also be potentially applied to other weights, though we did not observe extreme values for them.

Table 3 summarizes the 12 sets of different weights. The variability of the CCA-attr-w and ACA- attr-w weights based on baseline information is generally higher than that of the ACA-seq-attr-w weightings conditional on the previous wave. The sequential adjustment does not result in noisy weights because the wave-to-wave response rates are very high with large sample sizes. The maximum efficiency loss due to weighting, Loss(w), is defined as the squared value of the coefficient of variation (CV): $\text{Loss}(w) = cv^2(w) = \frac{var(w)}{mean^2(w)}$, and 1+ Loss(w) gives the design effects on the variance inflation due to weighting, which is calculated assuming independence between weights and the survey outcome variable (Kish, 1965). Little & Vartivarian (2005) recommend using mean squared error measures to reflect the weighting effect on the tradeoff between bias and variance, while 1+ Loss(w) serves the upper bound of the design effects. The efficiency loss is 33% for the CCA-attrition weight, around 28-29% for the sequentially adjusted ACA weights, and between 31–40% for the baseline adjusted ACA weights. Figure 1 shows that the collected outcome distributions do not substantially change across respondent groups defined by the weight quintiles, indicating that the outcome is weakly correlated with the weight, regardless of how it is calculated.

### *4.4.2 Imputation and sensitivity analysis*

We consider sequential MI (MI) and adding offsets to MI for sensitivity analysis (MI-offset). We conduct sequential conditional MI to handle both unit and item nonresponse using *mice* (Van Buuren, 2012), tailoring the imputation models to the incomplete variable types: logistic regression models for binary variables (e.g., sex and school types), multinomial logistic regression models for nominal variables (e.g., race/ethnicity), ordinal logistic regression models for ordered variables (e.g., poverty levels, social-economic status), tree-based methods for non-normally distributed numeric variables (e.g., age and percent of non-white students), and predictive mean matching for the assessment outcome (e.g., reading IRT scores). We run 15 or 30 iterations for each imputation during every wave and have conducted graphical diagnostics comparing imputed and observed values to assess convergence, where the number of iterations of sequential imputations is sufficient. We save five completed datasets for inferences after MI. The number of completed datasets can be substantially increased given sufficient computational resources. In our application we find that the impact on inferences is small, and five datasets are sufficient for NRBA.

*Table 4. Overall mean estimates (Est) of reading assessment scores with reported standard errors (SE) and bounds of the 95% confidence intervals (Lower, Up).*

|                | Est    | SE   | Lower   | Up     |
|----------------|--------|------|---------|--------|
| CCA            | 110.59 | 0.12 | 110.36  | 110.82 |
| CCA-base-w     | 110.39 | 0.13 | 110.14  | 110.64 |
| CCA-attr-w     | 110.08 | 0.13 | 109.82  | 110.34 |
| ACA            | 107.01 | 0.11 | 106.8   | 107.22 |
| ACA-base-w     | 106.76 | 0.12 | 106.53  | 107    |
| ACA-attr-w     | 106.69 | 0.12 | 106.44  | 106.93 |
| ACA-seq-attr-w | 106.66 | 0.12 | 106.42  | 106.9  |
| MI             | 109.56 | 0.13 | 109.3   | 109.83 |
| MI-offset(-0.8)| 108.53 | 0.14 | 108.23  | 108.83 |
| MI-offset(-1.2)| 107.93 | 0.11 | 107.71  | 108.14 |
| MI-offset(-1.6)| 107.33 | 0.11 | 107.12  | 107.55 |

*Note: CCA: complete case analysis; ACA: available case analysis; CCA-base-w: base weighted analysis accounting for the cluster structure by individuals with complete cases; ACA-base-w: base weighted analysis accounting for the cluster structure by individuals with available cases; CCA-attr-w: attrition-adjusted weighted analysis accounting for the cluster structure by individuals with complete cases; ACA-attr-w: attrition-adjusted, conditional on the baseline information, weighted analysis accounting for the cluster structure by individuals with available cases; ACA-seq-attr-w: attrition-adjusted, conditional on the previous information, weighted analysis accounting for the cluster structure by individuals with available cases; MI: multiple imputation with sequential wave by wave conditional imputation; MI-offset: sensitivity analysis with an offset $k_t \sigma_{race}$ added to the MI imputations, with different pre-specified values $k_t = -0.8, -1.2$ and $-1.6$; Single-imp: using the same specification model as MI but only conducting one imputation. SOURCE: U.S. Department of Education, National Center for Education Statistics, Early Childhood Longitudinal Study, Kindergarten Class of 2010-11, 2011-16 Spring.*

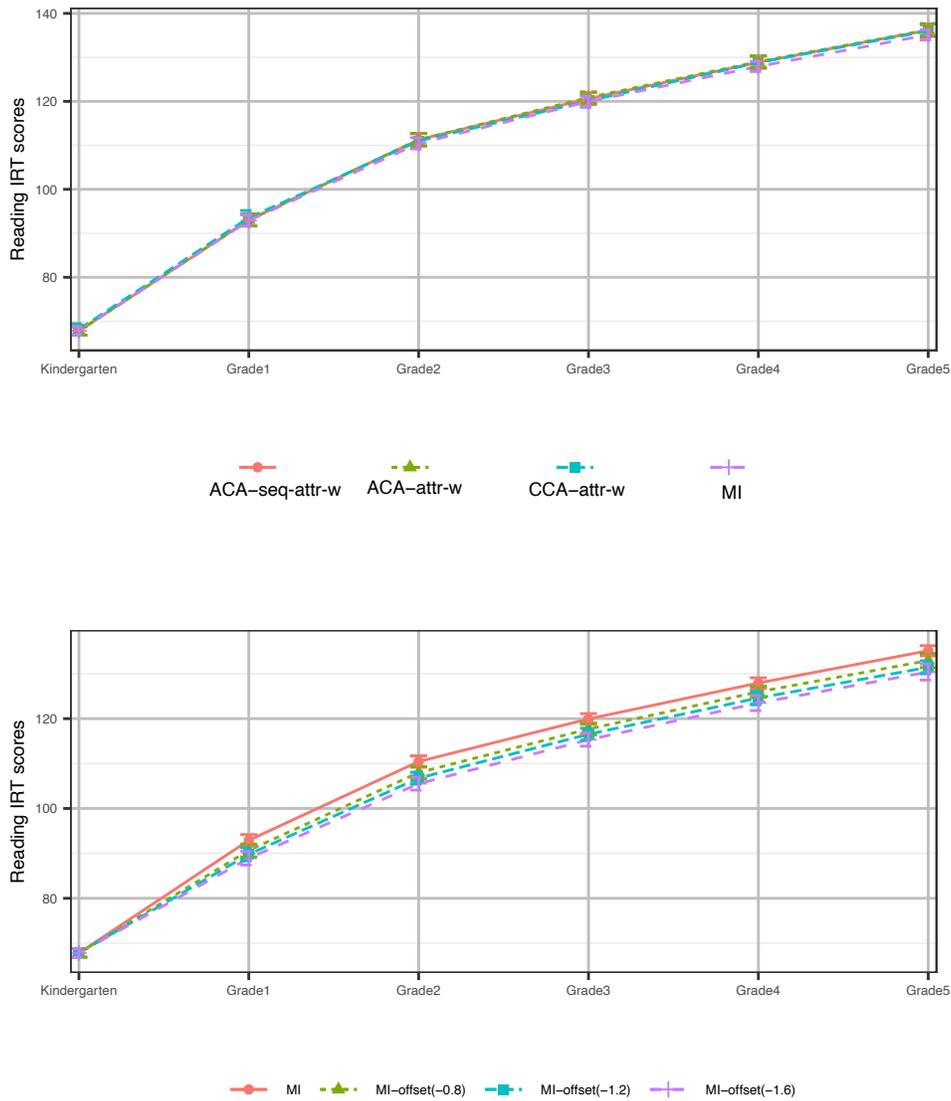

*Figure 2. Subgroup mean estimates of reading item response theory (IRT)-estimated scores for API students across grades (API: Asians, Native Hawaiians, and other Pacific Islanders; CCA-attr-w: attrition-adjusted weighted analysis accounting for the cluster structure by individuals with complete cases; ACA-attr-w: attrition-adjusted, conditional on the baseline information, weighted analysis accounting for the cluster structure by individuals with available cases; ACA-seq-attr-w: attrition-adjusted, conditional on the previous information, weighted analysis accounting for the cluster structure by individuals with available cases; MI: multiple imputation with sequential wave by wave conditional imputation; MI-offset: sensitivity analysis with an offset $k_t \sigma_{race}$ added to the MI imputations, with different pre-specified values $k_t = -0.8, -1.2$ and $-1.6$). SOURCE: U.S. Department of Education, National Center for Education Statistics, Early Childhood Longitudinal Study, Kindergarten Class of 2010-11, 2011-16 Spring.*

During the sensitivity analyses of MI-offset, we add the pre-described offset values to the missing values at the time of dropout and carry on the sequential imputation as conditional MAR. The choices of offset values are subjective and depends on the substantive questions. We assume that the attriters have lower assessment scores and use $k_t = -0.8, -1.2, -1.6$ as three examples following the practice of Giusti & Little (2011). The sensitivity parameter specification is subjective and depends on the substantive knowledge, and we recommend trying a wide range of values that may yield different findings.

*4.5* **Nonresponse bias analysis results**

We present estimates of the average assessment scores and regression coefficients of interest in a multilevel regression model. We compare the model with varying intercepts only across children and that with varying intercepts across both children and schools, and both models generate similar fixed-effect estimates.

The school-level variance is smaller than the child-level variance (~10%), so we omit school-varying intercepts for model simplicity. We calculate the weighted estimates and use the Taylor Series approximation for the variance estimation, implemented in the R package *survey* (Binder, 1983; Lumley, 2020). This also applies to the MI estimates when we account for the base weights to obtain the weighted estimate and within-imputation variance. Then we use MI combining rules to yield the total variance including the between-imputation variance.

Table 4 gives the overall mean estimates of the reading IRT scores in comparison of the approaches for descriptive summaries. Figure 2 gives the wave-specific averages for API children. Table 4 shows that both CCA and weighted CCA give the largest estimate, and the weights in CCA do not change the overall estimate. Including the available cases decreases the mean score, and the use of weights in ACA enlarges the decrease. MI yields values between the ACA and CCA estimates. The CCA, ACA, and MI analyses have different sample sizes.

Considering the monotone response pattern and increasing reading assessment scores across grades, ACA includes more observations with lower grades than CCA and thus decreases the overall mean score, and MI imputes the potential scores in higher grades for students who drop out and thus increases the mean value compared to ACA. Nevertheless, the differences are not substantial. Weights slightly change the estimates because they are weakly related to the outcome variable. ACA has lower variances than CCA mainly due to more included data. MI exploits covariates in the imputation model that are predictive of the missing values to reduce bias. The variances based on one randomly selected single imputation are mostly smaller than those of weighting, showing potential efficiency gains. As the combination of the average within imputation variances and between-imputation variance, MI variance accounts for the imputation model uncertainty and is comparable to the weighted estimates. When we add offsets in MI, the estimates decrease, indicating that the analysis is sensitive to deviations from MAR. When we add offsets in MI, Table 4 shows that the estimates significantly decrease from 109.56 with the 95% confidence interval (95% CI: 109.3-109.93) to 108.53 (95% CI: 108.23-108.83) with $k_t = -0.8$, 107.93 (95% CI: 107.71-108.14) with $k_t = -1.2$, and 107.33 (95% CI: 107.12-107.55) with $k_t = -1.6$, indicating that the overall mean estimates sensitive to deviations from MAR. Sensitivity analyses of the subgroup mean estimates show that with $k_t = -0.8$, the average reading scores of API students in Grades 1, 2, and 4 are significantly lower than the MI estimates, and their performances in Grades 3 and 5 substantially decrease when the offset is further reduced to -1.2. With $k_t = -1.6$, the reductions are enlarged.

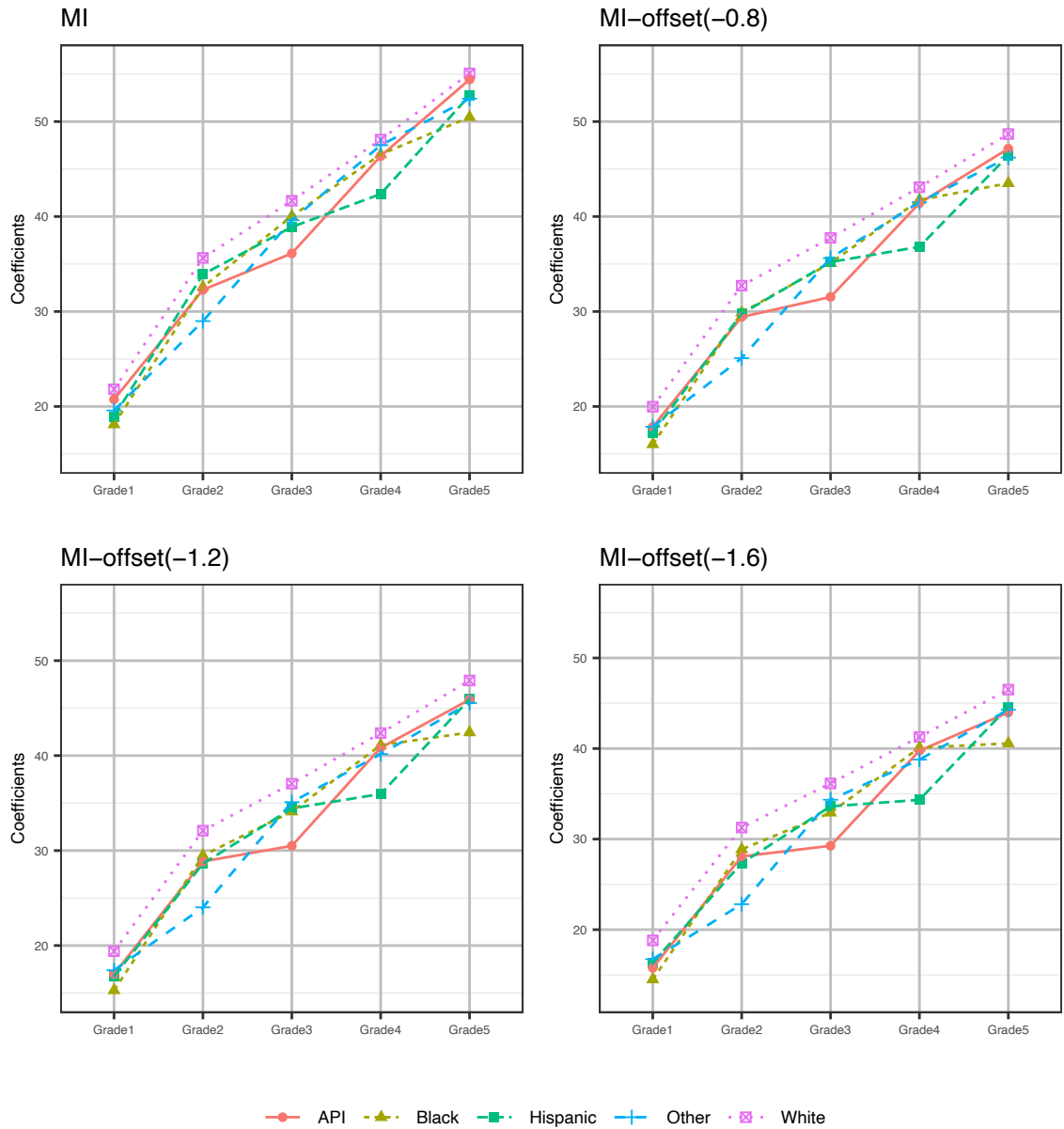

*Figure 3. Different wave effects on reading item response theory (IRT)-estimated scores across race/ethnicity groups based on multiple imputation and the inclusion of offsets (API: Asians, Native Hawaiians, and other Pacific Islanders; Non-Hisp: not Hispanic; Other: more than one race, not Hispanic; MI: multiple imputation with sequential wave by wave conditional imputation; MI-offset: sensitivity analysis with an offset $k_t \sigma_{race}$ added to the MI imputations, with different pre-specified values $k_t = -0.8, -1.2$ and $-1.6$). SOURCE: U.S. Department of Education, National Center for Education Statistics, Early Childhood Longitudinal Study, Kindergarten Class of 2010-11, 2011-16 Spring.*

Considering incomplete data modeling, we implement both multilevel models with random effects using ML estimation and marginal models using GEE. The multilevel model includes random intercepts $\alpha_i$ varying across children with a variance of $\sigma_0^2$ and random errors $\epsilon_{it}$ with a variance of $\sigma^2$: $\alpha_i \sim N(0, \sigma_0^2)$, $\epsilon_{it} \sim N(0, \sigma^2)$, which uses ML estimates for ACA and weighted ML estimates when incorporating weights into the pseudo-likelihood, respectively. The analysis model after MI is the multilevel model. We fit GEE with a pre-specified first-order autoregressive working correlation structure and weighted GEE to include the adjusted weights in the marginal model. We use large-sample approximations in the variance estimation, except for weighted GEE. The pseudo-ML and weighted GEE need to be combined with resampling methods to account for the sampling uncertainty.

Different weighting approaches have negligible effects on coefficient estimates, similar to the findings in Si (2022b). We compare the point estimates, standard error and significance levels based on the p-values. We declare that two estimates are similar if the significance conclusions are the same with slight differences in the magnitudes of the point and standard error estimates. We focus on different wave effects on reading IRT scores across race/ethnicity groups, i.e., the adjusted estimates of $(\vec{\beta}_8, \vec{\beta}_9)$ in Equation (6). The ML estimates are similar across the CCA, weighted CCA, ACA, and weighted ACA methods. The GEE and weighted GEE estimates are similar. GEE models use a different working correlation structure from the multilevel models, but the effect on coefficient estimates is small. The ML and GEE estimates are different from those under MI.

Adding offsets to MI changes the coefficient estimates, presented in Figure 3, which is as expected, since the offsets depend on residual standard deviations across race/ethnicity groups. The general differences across race/ethnicity groups are similar under different MI methods.

White students have more accelerated growth effects than students from other racial groups. But different approaches present different magnitudes of racial disparities. This shows that model fitting is sensitive to missing data mechanisms and not robust against model misspecification.

## 5  *Conclusions and discussion*

We compare weighting, MI, repeated measures models that allow for missing data, and sensitivity analysis approaches for NRBA in longitudinal studies. We consider complete cases, available cases, and completed data after imputation. Our NRBA focuses on the monotone response pattern, fits regressions of key survey outcomes and indicators of nonresponse on variables observed for both respondents and nonrespondents, compares estimates with and without nonresponse weighting adjustments, and implements sensitivity analyses based on adding offsets to MI results to assess the impact of deviations from MAR missingness. The NRBA of analytic inferences has different findings from that of descriptive summaries. Different adjustment approaches change descriptive summaries, and their effect on coefficient estimates depends on the model specification. All analyses can be carried out widely with available statistical software. Our example R code is available upon request.

The performance of methods analyzing incomplete longitudinal data depends on the missingness mechanism, and a challenge is that little is often known about this aspect. CCA is consistent for mechanisms that depend on baseline covariates but not the repeated measures. This includes MNAR mechanisms where missingness depends on covariates that have missing values. However, this method is not efficient if there is substantial information in the incomplete cases.

Weighted CCA can correct for bias in CCA when the mechanism is MAR but not MCAR, and the variables that define the weights are related to both the repeated measures and missingness. Weighted CCA is not recommended when the weights are unrelated to the repeated measures, since weighting then tends to increase variance with no compensating reduction in bias. The application of weighted ACA to monotone data by multiplying the conditional weights can weaken the MAR assumption by allowing missingness to depend on variables that are not fully observed over all waves. ML, GEE, and MI assuming MAR can handle general missingness patterns in the repeated measures, and MI can also handle missing data in baseline covariates. These methods can be more efficient than CCA and weighted CCA when the information in the incomplete cases is substantial for the analysis of interest. An advantage of MI over ML is that it allows variables to be included in the imputation model that are not included in the analysis model. ML is basically a large-sample technique, and Bayesian methods can provide better inferences than ML when sample sizes are small.

Comparisons of NRBA results from alternative methods for handling missing data provides an assessment of whether the choice of method has a small or large impact on findings. If results are sensitive to the choice of method, this suggests that nonresponse bias is more of a concern. If deviations from MAR are considered likely, a sensitivity analysis to assess the effect of deviations from MAR is suggested. We illustrate one such approach to this question with sensitivity analyses under MI to assess MNAR deviations and recommend more investigations.

Though modest differences are present for several estimates, we do not find substantial evidence of nonresponse bias in the ECLS-K:2011 study, perhaps reflecting the high response rates across waves conditional on the kindergarten response at the base year, omitting the baseline nonresponse, and the large proportions of movers who cannot continue the study, rather than attriters. However, lack of evidence of bias in the NRBA does not necessarily mean lack of bias; the key to a strong NRBA is the existence of a rich set of auxiliary variables that are highly predictive of the survey outcome variables. The auxiliary variables are collected from baseline or previous waves, across child, family, and school levels, available for both respondents and nonrespondents. The evidence is generally weak in this application because the observed covariates are not strongly related to the survey outcomes, where the assessment scores themselves are estimates based on the IRT models and subject to additional estimation error. NRBA findings for other survey variables may differ. We focus here on mean estimates for the population and population subgroups and regression models. In either setting, the key to a strong analysis is the availability of strong auxiliary variables that are not predictors in the regression model of interest.

In our NRBA application, we did not fit selection and pattern mixture models because they require strong but unverified assumptions, as a motivation for our sensitivity analysis. Our sequential MI does not explicitly include the two-level analysis model in the imputation specification but accounts for wave (and age) effects and the correlation of repeated measures by including previous measures as covariates in the sequential imputation model. However, the performance of the sequential MI is robust in the ECLS-K: 2011 study. We acknowledge the ideal property of this substantive model compatible MI that needs to accommodate practical implementation difficulties. More work is necessary to achieve successful implementation with large-scale application studies.

As regards future work, extension to multiple surveys with various quantities of interest could provide further evidence for the NRBA. Data integration of multiple sources can improve NRBA by providing better auxiliary information for nonresponse adjustment, and benchmark information for external validation. We focused on a monotone dropout, which is the predominant pattern in the ECLS-K:2011 study; MI is one approach that can handle intermittent missingness patterns (e.g., Si et al., 2020). Multiple missingness mechanisms are possible with different types of nonresponse. Hybrid approaches are also possible, such as using MI to create a monotone pattern and then sequentially weighting the resulting imputed data sets. Given rapidly decreasing response rates, probability samples become vulnerable to nonresponse bias. Assessing and adjusting for this bias requires an external reference sample of high quality or population distributions of highly predictive variables in the NRBA.

***References***
Andridge, R.R. & Little, R.J.A (2011). Proxy pattern-mixture analysis for survey nonresponse. *Journal of Official Statistics, 27*(2), 153-180.
Andridge, R.R. & Little, R.J.A. (2020). Proxy pattern-mixture analysis for a binary variable subject to nonresponse. *Journal of Official Statistics, 36*(3), 703–728.
Audigier, V., White, I. R., Jolani, S., Debray, T. P. A., Quartagno, M., Carpenter, J., Buuren, S. van, & Resche-Rigon, M. (2018). Multiple imputation for multilevel data with continuous and binary variables. *Statistical Science*, *33*(2), 160-183.
Bartlett, J. W., Seaman, S. R., White, I. R., & and, J. R. C. (2014). Multiple imputation of covariates by fully conditional specification: Accommodating the substantive model. *Statistical Methods in Medical Research*, *24*(4), 462–487.
Binder, D.A. (1983). On the variances of asymptotically normal estimators from complex surveys. *International Statistical Review, 51*(3), 279–292.
Breiman, L., Friedman, J.H., Olshen, R.A., & Stone, C.J. (1984). *Classification and Regression Trees,* Wadsworth, Belmont, CA, Chapman & Hall, New York.
Brick, J. M. (2013). Unit nonresponse and weighting adjustments: A Critical Review. *Journal of Official Statistics*, *29*(3), 329–353.
Brick, J. M., & Montaquila, J. M. (2009). Nonresponse and weighting. In *Handbook of statistics* (Vol. 29, pp. 163–185). Elsevier.
Brick, J. M., & Williams, D. (2013). Explaining rising nonresponse rates in cross-sectional surveys. *The ANNALS of the American Academy of Political and Social Science*, *645*, 36–59.
Buskirk, T.D., & Kolenikov, S. (2015). Finding respondents in the forest: A comparison of logistic regression and random forest models for response propensity weighting and stratification. *Survey Methods: Insights from the Field.* https://surveyinsights.org/?p=5108.
Büttner, T. J., Sakshaug, J. W., Vicari, B., & others. (2021). Evaluating the utility of linked administrative data for nonresponse bias adjustment in a piggyback longitudinal survey. *Journal of Official Statistics*, *37*(4), 837–864.

Cao, Y., Allore, H., Wyk, B. V., & Gutman, R. (2022). Review and evaluation of imputation methods for multivariate longitudinal data with mixed-type incomplete variables. *Statistics in Medicine, 41*(30), 5844-5876.

Chen, Q., Gelman, A., Tracy, M., Norris, F. H., & Galea, S. (2015). Incorporating the sampling design in weighting adjustments for panel attrition. *Statistics in Medicine*, *34*(28), 3637–3647.

Collins, L. M., Schafer, J. L., & Kam, C.-M. (2001). A comparison of inclusive and restrictive strategies in modern missing data procedures. *Psychological Methods*, *6*(4), 330-351.

Daniels, M. J., & Hogan, J. W. (2008). *Missing data in longitudinal studies: Strategies for Bayesian modeling and sensitivity analysis*. CRC Press.

de Leeuw, E., Hox, J., & Luiten, A. (2018). International nonresponse trends across countries and years: An analysis of 36 years of labor force survey data. *Survey Methods: Insights from the Field.* Retrieved from https://surveyinsights.org/?p=10452.

Demirtas, H. (2009). Multiple imputation for longitudinal data under a Bayesian multilevel model. *Communications in Statistics—Theory and Methods*, *38*(16–17), 2812–2828.

Dempster, A. P., Laird, N. M., & Rubin, D. B. (1977). Maximum likelihood from incomplete data via the EM algorithm (with discussion). *Journal of the Royal Statistical Society Series B*, *39*, 1–38.

Deng, Y., Hillygus, D. S., Reiter, J. P., Si, Y., & Zheng, S. (2013). Handling attrition in longitudinal studies: The case for refreshment samples. *Statistical Science*, *22*, 238–256.

Du, H., Enders, C., Keller, B. T., Bradbury, T. N., & Karney, B. R. (2021). A Bayesian latent variable selection model for nonignorable missingness. *Multivariate Behavioral Research*, *57*(2–3), 478–512.

Enders, C. K., Du, H., & Keller, B. T. (2020). A model-based imputation procedure for multilevel regression models with random coefficients, interaction effects, and nonlinear terms. *Psychological Methods*, *25*(1), 88–112.

Feldman, B.J., & Rabe-Hesketh, S. (2012). Modeling achievement trajectories when attrition is informative. *Journal of Educational and Behavioral Statistics*, *37*(6), 703–36.

Fitzmaurice, G., Davidian, M., Verbeke, G., & Molenberghs, G. (2009). *Longitudinal Data Analysis.* London: CRC Press.

Fitzmaurice, G. M., Laird, N. M., & Ware, J. H. (2012). *Applied longitudinal analysis*. Wiley.

Giusti, C., & Little, R.J.A. (2011). An analysis of nonignorable nonresponse to income in a survey with a rotating panel design, *Journal of Official Statistics, 27*(2), 211–229.

Graham, J. W., & Hofer, S. M. (2000). Multiple imputation in multivariate research. In *Modeling longitudinal and multilevel data* (pp. 189–204). Psychology Press.

Groves, R. M. (2006). Nonresponse rates and nonresponse bias in household surveys, *Public Opinion Quarterly, 70*(5), 2006, 646–675.

Grund, S., Lüdtke, O., & Robitzsch, A. (2016). Multiple imputation of missing covariate values in multilevel models with random slopes: A cautionary note. *Behavior Research Methods*, *48*(2), 640–649.

Grund, S., Lüdtke, O., & Robitzsch, A. (2019). Missing data in multilevel research, In S. E. Humphrey & J. M. LeBreton (Eds.), *The handbook of multilevel theory, measurement, and analysis* (pp. 365–386). American Psychological Association.

Hayes, T. (2019). Flexible, free software for multilevel multiple imputation: A review of Blimp and jomo. *Journal of Educational and Behavioral Statistics*, *44*(5), 625–641.

Haziza, D., & Lesage, É. (2016). A discussion of weighting procedures for unit nonresponse. *Journal of Official Statistics, 32*(1), 129-145.

He, Y., Zhang, G., & Hsu, C.-H. (2021). *Multiple imputation of missing data in practice: Basic theory and analysis strategies*. CRC Press.

Hedlin, D. (2020). Is there a `safe area' where the nonresponse rate has only a modest effect on bias despite non-ignorable nonresponse? *International Statistical Review*, *88*(3), 642–657.

Hothorn, T., Hornik, K., Strobl, C., & Zeileis, A. (2021). party: A laboratory for recursive partytioning. Version 1.3-6. http://cran.r-project.org/web/packages/party/.

Hothorn, T., Hornik, K., & Zeileis, A. (2006). Unbiased recursive partitioning: A conditional inference framework. *Journal of Computational and Graphical Statistics, 15*, 651-674.

Ibrahim, J. G., Chen, M. H., Lipsitz, S. R., & Herring, A. H. (2005). Missing data methods for generalized linear models: A comparative review. *Journal of the American Statistical Association*, *100*, 332–346.

Ibrahim, J. G., & Molenberghs, G. (2009). Missing data methods in longitudinal studies: A review. *Test*, *18*(1), 1–43.

Kalton, G., & Flores-Cervantes, I. (2003). Weighting methods. *Journal of Official Statistics*, *19*(2), 81–97.

Kaminska, O. (2022) Weighting panels together for cross-sectional estimation. *In Improving the Measurement of Poverty and Social Exclusion in Europe: Reducing Nonsampling Errors (ed.s P. Lynn & L. Lyberg), EU Publications.* Chapter 12, 173-183.

Keller, B., & Enders, C. (2021). Blimp user's guide version 3. *Los Angeles, CA*.

Kenward, M. G. (1998). Selection models for repeated measurements with non-random dropout: An illustration of sensitivity. *Statistics in Medicine*, *17*, 2723–2732.

Kenward, M. G., Molenberghs, G., & Thijs, H. (2003). Pattern-mixture models with proper time dependence. *Biometrika*, *90*(1), 53–71.

Kern, C., Weiß, B., & Kolb, J.-P. (2023). Predicting nonresponse in future waves of a probability-based mixed-mode panel with machine learning. *Journal of Survey Statistics and Methodology*, *11*(1), 100–123.

Kish, L, 1965. *Survey sampling*. New York, J. Wiley.

Liang, K.Y., & Zeger, S.L. (1986). Longitudinal data analysis using generalized linear models. *Biometrika, 7*, 13–22.

Little, R.J.A. (1993). Pattern-mixture models for multivariate incomplete data. *Journal of the American Statistical Association*, *88*(421), 125–134.

Little, R.J.A. (1994). A class of pattern-mixture models for normal incomplete data. *Biometrika*, *81*(3), 471–483.

Little, R.J.A., Carpenter, J., & Lee, K. (2022). A comparison of three popular methods for handling missing data: Complete-case analysis, weighting and multiple imputation. *Sociological Methods & Research, 0*(0). https://doi.org/10.1177/00491241221113873

Little, R.J.A., & David, M. (1983). Weighting adjustments for non-response in panel surveys. *Technical Report, US Dept. of Commerce, Bureau of the Census.*

Little, R.J.A., & Rubin, D.B. (2019). *Statistical Analysis with Missing Data,* 3rd Ed. Wiley: NY.

Little, R.J.A., & Vartivarian, S. (2003). On weighting the rates in non-response weights. *Statistics in Medicine, 22*(9), 1589-1599.

Little, R.J.A., & Vartivarian, S. (2005). Does weighting for nonresponse increase the variance of survey means? *Survey Methodology*, *31*, 161-168.

Little, R.J.A., & Wang, Y. (1996). Pattern-mixture models for multivariate incomplete data with covariates. *Biometrics*, *52*(1), 98–111.

Lohr, S., Hsu, V., & Montaquila, J. (2015). Using classification and regression trees to model survey nonresponse. *JSM Proceedings, Survey Research Methods Section*. Alexandria, VA: American Statistical Association. 2071–2085.
Lumley, T. (2020) survey: Analysis of complex survey samples. R package version 4.0.
Lüdtke, O., Robitzsch, A., & West, S. G. (2019). Analysis of interactions and nonlinear effects with missing data: A factored regression modeling approach using maximum likelihood estimation. *Multivariate Behavioral Research*, *55*(3), 361–381.
Lynn, P., & Watson, N. (2021) Issues in weighting for longitudinal surveys. *In Advances in Longitudinal Survey Methodology (ed. P. Lynn), Wiley*. Chapter 18, 447-468,
Muthén, B. (2004). Latent variable analysis: growth mixture modeling and related techniques for longitudinal data. *In: Kaplan D, editor. The SAGE handbook of quantitative methodology for the social sciences. Thousand Oaks: SAGE Publications.* 346–369.
Muthén, L.K., & Muthén, B.O. (2017). *Mplus User's Guide. Eighth Edition*. Los Angeles, CA.
Natarajan, S., Lipsitz, S., Fitzmaurice, G., Moore, C., & Gonin, R. (2008). Variance estimation in complex survey sampling for generalized linear models. *Journal of the Royal Statistical Society. Series C (Applied Statistics)*, *57*(1), 75-87.
Pfeffermann, D., Skinner, C. J., Holmes, D. J., Goldstein, H., & Rasbash, J. (1998). Weighting for unequal selection probabilities in multilevel models. *Journal of the Royal Statistical Society Series B Stat. Methodol*, *60*(1), 23–40.
Quartagno, M., & Carpenter, J. (2022). Substantive model compatible multilevel multiple imputation: A joint modeling approach. *Statistics in Medicine*, *41*(25), 5000–5015.
Quartagno, M., Grund, S., & Carpenter, J. (2019). Jomo: A flexible package for two-level joint modelling multiple imputation. *R Journal*, *11*(2), 205-228.
Rabe-Hesketh, S., & Skrondal, A. (2006). Multilevel modelling of complex survey data. *Journal of the Royal Statistical Society Series A*, *169*(4), 805–827.
Raghunathan, T., Lepkowski, J. VanHoewyk, M., & Solenberger, P. (2001). A multivariate technique for multiply imputing missing values using a sequence of regression models, *Survey Methodology.* 27(1), 85-95.
Robitzsch, A., & Lüdtke, O. (2021). *mdmb*: Model based treatment of missing data. R package.
Robins, J.M., Rotnitzky, A., & Zhao, L.P. (1995). Analysis of semiparametric regression models for repeated outcomes in the presence of missing data. *Journal of the American Statistical Association*, *90*(429), 106-121.
Rubin, D.B. (1976). Inference and missing data. *Biometrika, 63*(3), 581-592.
Rubin, D.B. (1987). *Multiple Imputation for Nonresponse in Surveys*. NY: John Wiley & Sons.
Sánchez-Cantalejo, C. et al. (2021). Impact of COVID-19 on the health of the general and more vulnerable population and its determinants: Health Care and Social Survey–ESSOC, Study Protocol. *Int. J. Environ. Res. Public Health 18*, 8120.
Schafer, J. L. (1997). *Analysis of Incomplete Multivariate Data*. London: Chapman & Hall.
Scharfstein, D. O., Rotnitzky, A., & Robins, J. M. (1999). Adjusting for nonignorable drop-out using semiparametric nonresponse models. *Journal of the American Statistical Association*, *94*(448), 1096–1120.
Seaman, S. R., Bartlett, J. W., & White, I. R. (2012). Multiple imputation of missing covariates with non-linear effects and interactions: An evaluation of statistical methods. *BMC Medical Research Methodology*, *12*(1), 1–13.

Si, Y. (2012). *Nonparametric Bayesian methods for multiple imputation of large scale incomplete categorical data in panel studies* [PhD Thesis]. Duke University.
Si, Y., Little, R.J.A., Mo, Y., & Sedransk, N. (2022a). A Case Study of Nonresponse Bias Analysis in Educational Assessment Surveys. *Journal of Educational and Behavioral Statistics*, *48*(3), 271-295.
Si, Y., Palta, M., & Smith, M. (2020). Bayesian profiling multiple imputation for missing hemoglobin values in electronic health records. *Annals of Applied Statistics, 14*(4), 1903–1924.
Si, Y. & Reiter, J.P. (2013), Nonparametric Bayesian multiple imputation for incomplete categorical variables in large-scale assessment surveys. *Journal of Educational and Behavioral Statistics, 38*(5), 499–521.
Si, Y., Reiter, J.P., & Hillygus, D.S. (2015). Semi-Parametric selection models for potentially non-ignorable attrition in panel study with refreshment samples. *Political Analysis, 23*, 92–112.
Si, Y., Reiter, J.P., & Hillygus, D.S. (2016). Bayesian latent pattern mixture models for handling attrition in panel studies with refreshment samples. *The Annals of Applied Statistics, 10*, 118–43.
Si, Y., West, B. T., Veliz, P., Patrick, M. E., Schulenberg, J. E., Kloska, D. D., Terry-McElrath, Y. M., & McCabe, S. E. (2022b). An empirical evaluation of alternative approaches to adjusting for attrition when analyzing longitudinal survey data on young adults' substance use trajectories. *International Journal of Methods in Psychiatric Research*, *31*(3), e1916, <https://doi.org/10.1002/mpr.1916>.
Valliant, R., Dever, J.A., & Kreuter, F. (2018) *Practical Tools for Designing and Weighting Survey Samples.* 2nd ed. New York: Springer.
Van Buuren, S. (2011). Multiple imputation of multilevel data. In *Handbook of advanced multilevel analysis* (pp. 181–204). Routledge.
Van Buuren, S. (2012), *Flexible Imputation of Missing Data*, CRC Press.
Wager, S., Hastie, T. & Efron, B. (2014). Confidence intervals for random forests: The jackknife and the infinitesimal jackknife, *The Journal of Machine Learning Research 15*(1), 1625–1651.
Wu, M. C., & Carroll, R. J. (1988). Estimation and comparison of changes in the presence of informative right censoring by modeling the censoring process. *Biometrics*, *44*(1), 175–188.
Xie, X., & Meng, X. (2017), Dissecting multiple imputation from a multi-phase inference perspective: What happens when God's, Imputer's and Analyst's models are uncongenial? *Statistica Sinica*, *27*(4), 1485-545.
Yucel, R. M. (2008). Multiple imputation inference for multivariate multilevel continuous data with ignorable non-response. *Philosophical Transactions of the Royal Society A: Mathematical, Physical and Engineering Sciences*, *366*(1874), 2389–2403.
Zhou, M., & Kim, J. K. (2012). An efficient method of estimation for longitudinal surveys with monotone missing data. *Biometrika, 99*, 631–648.